\documentclass[preprintnumbers,superscriptaddress]{revtex4}
\usepackage{amsfonts}
\usepackage{amssymb}
\usepackage{amsmath}
\usepackage{epsfig}
\usepackage{float}
\usepackage{subfig}
\usepackage{epstopdf}
\usepackage{graphicx}
\usepackage{array}
\usepackage{booktabs}
\usepackage{mathrsfs}
\usepackage{dsfont}
\usepackage{xcolor}

\allowdisplaybreaks

\begin{document}
\makeatletter
\def\@pacs{}
\makeatother


\title{Completely Deformed Complexity-free Anisotropic Fluid Spheres}
\author{Z. Yousaf}
 \email{zeeshan.math@pu.edu.pk}
\affiliation{Department of Mathematics, University of the Punjab, Quaid-i-Azam Campus, Lahore-54590, Pakistan.}


\author{Kazuharu Bamba}
 \email{bamba@sss.fukushima-u.ac.jp}
\affiliation{Faculty of Symbiotic Systems Science,
Fukushima University, Fukushima 960-1296, Japan.}%


\author{M. Z. Bhatti}
\email{mzaeem.math@pu.edu.pk}
\affiliation{Department of Mathematics, University of the Punjab, Quaid-i-Azam Campus, Lahore-54590, Pakistan.}
\affiliation{Research Center of Astrophysics and Cosmology, Khazar University, 41 Mehseti Street, 1096, Baku, Azerbaijan.}

\author{S. Khan}
\email{suraj.pu.edu.pk@gmail.com; suraj.math@uaf.edu.pk}
\affiliation{University of Agriculture Faisalabad, Constituent
College, Toba Tek Singh 36050, Pakistan.}

\date{\today}

\begin{abstract}

This investigation highlights an important application of complete geometric decoupling in constructing anisotropic, compact density-matter stars via a decoupled gravitational framework. In this context, the present study introduces an intriguing synthesis of two independent techniques, density-like constraints and the zero-complexity factor, to simultaneously derive the decoupler functions. By using this innovative relativistic scheme, we gain the ability to analytically control the anisotropies and complexity when modeling the dense-matter compact stars. We show that the complexity-free condition effectively captures the influence of anisotropic pressure inherent in compact dense-matter distributions, arising naturally from the chosen seed metric ansatz. Two distinct and physically viable anisotropic models satisfying all standard stability and energy conditions are obtained through the complete decoupling process. Our findings provide clear theoretical understanding of the coupling between known and standard gravity fields by demonstrating for the first time that the parameter responsible for deformation uniquely governs the direction of energy transfer between the seed sector and the decoupling source.

\end{abstract}
\maketitle
\onecolumngrid


\section{\label{sec:level1}Introduction}


Exploring the behavior of compact density-matter distributions is vital for a deeper investigation and validation of Einstein’s general relativity (GR). Black holes (BHs) represent the most remarkable ultra-compact matter distributions and serve as ideal candidates for exploring intense gravitational phenomena in the cosmos. This recognition is largely attributed to their distinctive properties, such as unstable photon spheres. Astrophysicists analyze gravitational waves (GWs) caused by the collisions of BHs, neutron stars, and other ultra-compact celestial remnants \cite{barack2019black}. The findings of GR \cite{cardoso2009geodesic} suggest that the quasinormal modes of GWs correspond to the pathways of light surrounding merging objects. This straightforward relationship is not followed by all gravity theories \cite{konoplya2017eikonal}. Even GR can break when non-linear electrodynamics are taken into account. Unveiling the mysteries associated with interior compositions of self-gravitational configurations is important, particularly for a deep grasp of the relativistic gravitational collapse of compact objects. Stellar interiors are believed to consist of a complex system of different fluids, interacting through non-linear processes. One of the simplest models used to approximate the interior of self-gravitational compact objects is based on the equality of principle stresses, i.e., $P=P_{r}=P_{\bot}$ \cite{boonserm2005generating}. However, this assumption significantly reduces the complexity of constructing the dense-matter configurations by solving the GR-field equations. Although ideal fluid models simplify the study of gravitationally bound stellar spheres, some relativistic models based on this assumption do not represent a physically acceptable situation, as indicated by the analysis of Lake and Delgaty \cite{delgaty1998physical}. Popławski \cite{poplawski2013intrinsic} demonstrated that, in curved spacetime, the intrinsic angular momentum of matter necessitates the metric-affine formulation of gravity, in which torsion is treated as a dynamical variable. This formulation generalizes general relativity into the simplest gravitational theory that incorporates intrinsic spin, known as the Einstein–Cartan–Sciama–Kibble theory. Grezia \emph{et al.} \cite{di2017spin} explored the construction of spinning matter-regulated traversible wormholes with spherical fluid configuration, by employing the Einstein-Cartan gravity model. Astashenok and Odintsov \cite{astashenok2020supermassive} analyzed realistic neutron stellar configurations in $R^2$ gravity. They identified that the scalar curvature drops more slowly because of an axion envelope around the configuration. Furthermore, Popławski \cite{poplawski2021gravitational} studied the gravitational collapse of a spherically symmetric spin fluid using the Tolman metric and the Einstein-Cartan theory. The author showed that the quantum particle production amplifies torsion effects, triggering inflation and the creation of matter. Several bounces may occur in the resulting closed universe, suggesting that the cosmos could have originated from a black hole. The insights given in \cite{poplawski2012four,poplawski2013cosmological,khlopov1985gravitational,de2020general,de2021testing,nojiri2020f,khan2025modeling} must be taken into consideration to attain a broader perspective of gravitational phenomena and their involvement in the genesis of stellar structures.

The extensive study of ultra-dense self-gravitational configurations has allowed us to gain a fundamental understanding of the composition and chronological history of the cosmos. By serving as natural laboratories, these enormous cosmic structures enable astronomers to conduct extensive surveys such as the Two-degree Field Galaxy Redshift Survey, the Large Synoptic Survey Telescope, and the Sloan Digital Sky Survey, which help uncover the mysteries of the universe. It has been a fascinating task for researchers to develop a universal metric for measuring the complexity factor (CF, thereafter) of various structures across multiple disciplines \cite{kolmogorov1965three,calbet2001tendency,sanudo2008statistical,panos2009simple}. The definition of CF might vary significantly depending on the situation and the type of physical problem. One example of a system with zero CF is a perfect crystal, which has a highly organized and symmetrical composition. The atomic structure of an isolated ideal gas, on the other hand, shows a sharp contrast to a symmetrical crystal configuration. Because of its total disorder, its constituent parts have an equal probability of occupying any condition that is accessible. This leads to a maximally complex system that is hard to predict because it encodes a large amount of information in its non-symmetric shape. However, the completely symmetrical crystal and the non-symmetric ideal gas display different behaviors within the framework of the idea of disequilibrium. Owing to its non-periodic nature, the
ideal gas displays minimal disequilibrium. In contrast, the perfect
crystal, characterized by its periodic configuration, manifests
maximum disequilibrium. Thus, the contrasting roles of information
and disequilibrium in complexity may be captured by characterizing CF as
their combined effect, an outcome of these notions. Therefore, the
definition is consistent with the intuitive concept of minimal CF
for both the perfect crystal and ideal gas.

The inclusion of pressure anisotropy, resulting from unequal principal stresses denoted as $\Delta\equiv P_{r}-P_{\bot}$, has a long and distinguished history in exploring the dynamics of compact stars \cite{bowers1974anisotropic,heintzmann1975neutron,cosenza1982evolution,herrera1985isotropic,
ponce1987general,herrera1997local,dev2002anisotropic,mak2002exact,mak2003anisotropic,
negreiros2009electrically,bhatti2021electromagnetic,yousaf2022f,khan2024structure}. The behavior of the stellar pressure may not remain the same within dense-matter stellar formations due to extreme density and the dominance of the gravitational field. These configurations feature non-uniformity of the fluid pressure, which leads to the concept of radial and tangential stresses. These types of compact stars could be valuable tools for exploring phase transitions and the interiors of stellar structures characterized by the combination of two perfect fluids.  In this direction, Letelier \cite{letelier1980anisotropic} proposed a two-perfect-fluid model of an anisotropic fluid in which the seed stellar fluid can be expressed as the sum of the stress-energy tensors of two independent perfect fluids. Pioneering research by Lema\^{i}tre \cite{lemaitre1933univers} emphasized the significance of this result for the structure and evolution of compact stellar configurations. Afterward, the investigations by Bowers and Liang \cite{bowers1974anisotropic} brought renewed attention to the fascinating topic of anisotropic relativistic matter distributions within the arena of Einstein's GR. By extending the general relativistic conservation equation, they constructed a time-independent self-gravitational stellar fluid endowed with spherical symmetry. They also analyzed the variations in the gravitational mass and surface redshift. The theoretical investigations carried out by Ruderman \cite{ruderman1972pulsars} showed that the nuclear matter exhibits anisotropic features at extremely high densities, i.e., of the order of $10^{15} \textmd{g}/\textmd{cm}^3$. Ruderman's work showed that the pressure splits into $P_{r}$ and $P_{\bot}$ in highly compact stellar distributions. In this context, several physical processes have been proposed as potential candidates for the existence of anisotropy in compact self-gravitational configurations. These include different types of phase transitions \cite{sokolov1980phase}, the existence of a solid core, the sum of two fluid configurations, the occurrence of type 3A superfluid \cite{kippenhahn1990stellar}, and other physical factors. In recent years, an increasing amount of research has explored anisotropic stellar models in a variety of contexts \cite{yousaf2022analysis,khan2024complexity,albalahi2024electromagnetic,yousaf2024modeling,khan2025einasto,khan2025zhao}.
The inclusion of anisotropic pressure for the exploration of compact stellar configuration, whether Newtonian or relativistic, is essential because it occurs in many realistic scenarios (see \cite{herrera1997local} for details). Furthermore, the authors of \cite{herrera2020stability} further shows that stellar evolution naturally leads to pressure anisotropy, even if the initial state is isotropic. The key takeaway is that any equilibrium configuration is the endpoint of a dynamic process. Therefore, it is logical to expect that the stellar configuration will maintain its anisotropic properties following this process. Hence, although a system starts with isotropic pressure, the resulting equilibrium configuration should ideally become anisotropic. Apart form GR, recent applications of modified $f(R)$ gravity also highlight the strong connection between black holes and cosmic dark matter. In this context, Khan and Yousaf \cite{khan2024construction} successfully developed fuzzy matter distributions for dark matter-based black holes within the framework of the exponential $f(R)$ model.

Additionally, attempts have also been undertaken by employing the principles of GR to define the CF for dense-matter stellar configurations. Notably, these approaches possess deficiencies that require attention \cite{panos2009simple,sanudo2009complexity,chatzisavvas2009complexity,de2012entropy,de2014information}. These studies explore the complexity of stellar fluids solely in terms of their energy density. However, this definition appears incomplete because it omits pressure, a crucial component of stellar fluids. Working in this direction, Herrera extended the CF by incorporating pressure anisotropy alongside energy density. This allowed him to analyze the CF for static/non-static anisotropic stellar formations within the framework of GR \cite{herrera2018new,herrera2018definition}. This extended version of the CF primarily originates from splitting
the Riemann tensor orthogonally. This splitting, in turn, provides a
structure scalar ($Y_{\textmd{TF}}$) referred to as the CF. The
reason for nominating $Y_{\textmd{TF}}$ as the CF is
because it relates the anisotropic stresses, density gradient, and
the Tolman mass in a specific for static, highly dense stellar
configurations. This definition has been built on the assumption
that a complexity-free, static stellar distribution is
homogeneous in energy density and isotropic in terms of pressure.
This implies that a stellar fluid sphere with a zero CF is
characterized by either; i) constant density and
pressure isotropy, or ii) non-uniformness of density and pressure anisotropy that
somehow cancel each other out to produce a zero net effect on the
CF. Several researchers have employed the zero CF condition as an
additional constraint to close the system of stellar structure equations
characterizing the dynamics of anisotropic stellar fluid spheres. By making use of this alternative description of the CF, researchers have formulated several analytical stellar models that describe the evolution of dissipative anisotropic stellar distributions within various gravitational models \cite{yousaf2022stability,khan2025modeling}. In the theory of superdense stellar compositions, this framework inherently leads to a definition of a system with zero CF, thereby establishing the concept of complexity. Furthermore, the CF enables us to identify a family of stellar models that share the same level of complexity. Within this context, it is especially intriguing to probe the implications of zero CF on the inherent structural attributes of the stellar interiors.

We can assign a specific value to the complexity-determining scalar function
$Y_{\textmd{TF}}=0$, indicating a system with no complexity. In this case, it functions like an equation of state (EoS), enabling us to construct the solution of stellar structure equations. However, obtaining analytical solutions for even such a simplified system can still be extremely challenging, if not impossible. To tackle the complexities arising from finding the solutions of GR field equations in an anisotropic domain, gravitational decoupling (GD) \cite{ovalle2013role,ovalle2017decoupling,ovalle2018black,ovalle2019decoupling} is considered an effective ingredient. The scheme of GD can be applied within stellar systems in two ways: the minimal geometric deformation (MGD) (the simplest form of GD), and the extended MGD alternatively known as the complete geometric deformation scheme.
Driven by the recent surge of interest in anisotropic stresses, this paper applies the CGD approach to decouple gravitational sources incorporating the additional term $\Theta_{\mu\nu}$.
This method utilizes known solutions as a foundation leading to the development of new ones and simplifies the problem by reducing it to a system of less complex differential equations. Furthermore, we can illustrate the CGD scheme through a specific example. Consider two gravitational-field sources, $\mathcal{G}_{1}$ and $\mathcal{G}_{2}$, where $\mathcal{G}_{1}$ is associated with standard GR-field equations, while $\mathcal{G}_{2}$ corresponds to the quasi-Einstein gravitational system. We solve each of the two field sources separately after defining each of them. The complete model for the system is finally found by joining the two solutions, and it is denoted as $\mathcal{G}_{1}\cup\mathcal{G}_{2}$. Furthermore, it is widely recognized that GR-field equations exhibit a highly nonlinear nature. Hence, the CGD approach stands out as a potent methodology for addressing them. The authors of \cite{misyura2024non} describe the effects of GD on regular BHs within the background of static, self-gravitational sources. Additionally, the influence of geometrical deformations on BH shadow in general stationary and static cases can be found in the studies \cite{vertogradov2024analyzing,vertogradov2024analyzing}, and \cite{vertogradov2024influence} explore the influence of a primary hair on a photon sphere, shadow and intensity distribution. In addition, some interesting analytical solutions characterizing the self-gravitational compact configurations have been presented in \cite{heydarzade2023hairy,vertogradov2022vaidya}.

This work is concerned with exploring the viable construction of anisotropic interior stellar models by making use of the CGD scheme, which is known as the extended form of the MGD decoupling. The MGD method transformed the known seed-gravitational system into an anisotropic regime by deforming the radial metric component only. In this approach, the energy exchange between the relativistic fluids is zero. However, the CGD approach enables us to obtain the anisotropic stellar models by deforming the radial as well as temporal metric potentials. In this context, we explore the transfer of energy between the seed-gravitational field and the generic field source $\Theta_{\mu\nu}$. This work extends our previous investigation \cite{yousaf2024modelingpress} by utilizing either the extended MGD scheme or the CGD decoupling. The previous investigation was grounded in the MGD method of decoupling gravitational sources.
 The remainder of this paper is structured as follows: Section \textbf{II} introduces the essential ingredients of the gravitational complexity approach within a spherically symmetric dense-matter system sourced by two fluids. In Section \textbf{III}, we recall the conventional method of gravitational decoupling, emphasizing its role in describing anisotropic fluid spheres and their internal dynamics. Section \textbf{IV} develops fully deformed Kohler–Chao–Tikekar models, rendered complexity-free through the imposition of a density-like condition in the anisotropic setting. Section \textbf{V} then addresses the physical analysis of the proposed models, while Section \textbf{VI} concludes with a discussion of the main outcomes of this work.

\section{The Notion of Gravitational Complexity}

In this section, we review the pioneering studies by Herrera and collaborators on the behavior of anisotropic stellar spheres revealing that the mechanism and evolution of these highly dense configurations can be determined using some specific scalar terms \cite{herrera2004spherically,herrera2009structure}. These scalar terms are the primary components of some tensorial quantities and may be written down with the help of the Riemann tensor $R_{\mu\nu\omega\varepsilon}$ as
\begin{align}\label{z1}
Y_{\mu\nu}&=R_{\mu\varepsilon\nu\omega}U^{\varepsilon}U^{\omega},
\\\label{z2}
Z_{\mu\nu}&=^{\ast}R_{\mu\varepsilon\nu\omega}U^{\varepsilon}U^{\omega},
\\\label{z3}
X_{\mu\nu}&=^{\ast}R^{\ast}_{\mu\varepsilon\nu\omega}U^{\varepsilon}U^{\omega},
\end{align}
which is known as orthogonal splitting of $R_{\mu\nu\varepsilon\epsilon}$ \cite{bel1961inductions}. Here, $U^{\mu}$ denotes the four-velocity. The symbol $\ast$ signifies the dual tensor, i.e., $\frac{1}{2}\eta_{\epsilon\gamma\varepsilon\omega}R^{~~~\epsilon\gamma}_{\mu\nu}$, with $\eta_{\epsilon\gamma\varepsilon\omega}$ being the Levi-Civita tensor.
Among the above-stated quantities, $Y_{\mu\nu}$ is particularly significant as the subject of this study, since the variable suppose to determine the CF of anisotropic matter spheres is associated with it. Being the central part of Herrera's ground-breaking notion of gravitational complexity to analyze the physical properties of the anisotropic stellar fluids \cite{herrera2018new}, $Y_{\mu\nu}$ can be expressed as a sum of its trace and trace-free parts as \cite{herrera2009structure}
\begin{align}\label{z4}
Y_{\mu\nu}&=Y_{<\mu\nu>}+\frac{1}{3}Tr(Y)
h_{\mu\nu},
\end{align}
where
\begin{align}\label{z5}
Y_{<\mu\nu>}=\left(s_{\mu}s_{\nu}-\frac{1}{3}h_{\mu\nu}\right)Y_{\mathrm{TF}};\quad  Tr(Y)=Y^{\mu}_{\mu}\equiv Y_{\mathrm{T}},
\end{align}
where $s^{\mu}$ and $h^{\mu}_{\nu}$ are defined by
\begin{align}\label{z6}
s^{\mu}=(0,e^{-\beta/2},0,0);\quad h^{\mu}_{\nu}=\delta^{\mu}_{\nu}-U^{\mu}U_{\nu},
\end{align}
such that $U_{\mu}s^{\mu}=0,~s_{\mu}s^{\mu}=-1$.
Later on, we will see that $Y_{TF}$ plays a crucial role in determining the local anisotropy of the self-gravitational compact configuration.
\begin{align}\label{z7}
ds^{2}=\textsl{g}_{\mu\nu}dx^{\mu}dx^{\nu}=e^{\alpha(r)}dt^{2}-e^{\beta(r)}dr^{2}-r^{2}\left(d\theta^{2}+\sin^{2}\theta
d\phi^{2}\right),
\end{align}
where the functions $\alpha(r)$ and $\beta(r)$ are the gravitational potentials. The choice of this ansatz is physically motivated, as it represents the most general line element for a non-rotating, spherically symmetric compact configuration. The temporal component encodes the redshift of signals due to gravitational effects, while the radial component reflects the spatial curvature produced by matter and energy. The angular sector describes the geometry of a $2$-sphere, consistent with spherical symmetry. This ansatz is therefore well-suited for analyzing relativistic compact stars and other static self-gravitating configurations. The metric \eqref{z7} must satisfy the GR-field equations
\begin{align}\label{z8}
G^{\mu}_{\nu}
=8\pi T^{\mu}_{\nu};\quad T^{\mu}_{\nu}=\textmd{dia\textsl{g}}\left(\sigma,-P_{r},
-P_{\bot},-P_{\bot}\right),
\end{align}
where the radial and tangential stresses are denoted by $P_{r}$ and $P_{\bot}$, respectively, while $\sigma$ denotes the stellar structure's energy density. The non-zero components of Eq. \eqref{z8} reads
\begin{align}\label{z9}
8\pi T^{0}_{0}&=
\frac{1}{r^{2}}+e^{-\beta}\left(\frac{\beta'}{r}-\frac{1}{r^{2}}\right),
\\\label{z10}
8\pi T^{1}_{1}&=
-\frac{1}{r^{2}}+e^{-\beta}\left(\frac{\alpha'}{r}+\frac{1}{r^{2}}\right),
\\\label{z11}
8\pi
T^{2}_{2}&=8\pi
T^{3}_{3}=\frac{e^{-\beta}}{4}\left(2\alpha''+\alpha'^{2}-\beta'\alpha'+2\frac{\alpha'}{r}-2\frac{\beta'}{r}\right).
\end{align}
Since the stress-energy tensor satisfies the relation $\nabla^{\mu}T_{\mu\nu}=0$, which gives
\begin{align}\label{z12}
\left(T^{1}_{1}\right)'-\frac{\alpha'}{2}\left(T^{0}_{0}-T^{1}_{1}\right)-\frac{2}{r}\left(T^{2}_{2}-T^{1}_{1}\right)=0,
\end{align}
with
\begin{align}\label{z13}
\alpha'=\frac{2m+8\pi r^{3}P_{r}}{(r-2m)}.
\end{align}
Next, the formula encoding the mass of spherical stellar fluids according to the Misner and Sharp formalism reads \cite{misner1964relativistic}
\begin{align}\label{z14}
R^{3}_{232}=1-e^{-\beta}=\frac{2m}{r},
\end{align}
or
\begin{align}\label{z15}
m(r)=4\pi\int^{r}_{0}t^{2}T^{0}_{0}(t)dt,
\end{align}
which can be expressed in a specific form that captures the contributions of both uniform and non-uniform density as
\begin{align}\label{z16}
m=\frac{4\pi}{3}r^{3}T^{0}_{0}(r)-\frac{4\pi}{3}\int^{r}_{0}t^{3}\left[T^{0}_{0}(t)\right]'dt.
\end{align}
Consequently, the trace-free component of $Y_{\mu\nu}$ capturing the complexity of static, self-gravitational stellar configurations with anisotropic stresses takes the form
\begin{align}\label{z17}
Y_{\mathrm{TF}}=8\pi(T^{2}_{2}-T^{1}_{1})-\frac{4\pi}{r^{3}}\int^{r}_{0}t^{3}\left[T^{0}_{0}(t)\right]'dt,
\end{align}
where $\Delta\equiv T^{2}_{2}-T^{1}_{1}$. This concept is rooted in the principle that the complexity-free stellar configurations satisfy the constraint $\Delta=0\Rightarrow T^{1}_{1}=T^{2}_{2}$. Additionally, $Y_{\mathrm{TF}}$ signifies how the Tolman mass $m_{\mathrm{T}}$ is effected by combined effect of  anisotropic factor $\Delta(r)$ and non-uniformity of density. The symbol $m_{\mathrm{T}}$ has the following representation for the considered source
\begin{align}\label{z18}
m_{\mathrm{T}}=4\pi\int^{r}_{0}t^{2}e^{(\alpha+\beta)/2}\left(\sigma+P_{r}+2P_{\bot}\right)dt,
\end{align}
which is a measure of total energy within an anisotropic spherical stellar structure. The Tolman mass can be represented in term of $m(r)$ with the help of following expression \cite{herrera1998role}
\begin{align}\label{z19}
m_{\mathrm{T}}=e^{(\alpha-\beta)/2}\left[m(r)-4\pi r^{3}T^{1}_{1}\right],
\end{align}
which by using the values of $m$ and $T_{1}^{1}$ takes the form
\begin{align}\label{z20}
m_{\mathrm{T}}=\frac{\alpha'}{2}r^{2}e^{(\alpha-\beta)/2}.
\end{align}
The term $Y_{\mathrm{TF}}$ allows us to calculate the mass function, $m_{\mathrm{T}}$ as
\begin{align}\label{z21}
m_{\mathrm{T}}=M_{\mathrm{T}}\left(\frac{r}{R}\right)^{3}+r^{3}\int^{r}_{0}\frac{e^{(\alpha+\beta)/2}}{t}
Y_{\mathrm{TF}}dt,
\end{align}
where $M_{\mathrm{T}}$ corresponds to the total value of $m_{T}$ for a spherical stellar source of radius $R$. Next, the zero CF condition reads
\begin{align}\label{z22}
Y_{\mathrm{TF}}=0\Rightarrow \Delta=(T^{2}_{2}-T^{1}_{1})=\frac{1}{{2}r^{3}}\int^{r}_{0}t^{3}\left[T^{0}_{0}(t)\right]'dt.
\end{align}
Then, by using the GR-field equations \eqref{z9}--\eqref{z11}, the scalar $T_{\mathrm{TF}}$ takes the form
\begin{align}\label{z23}
Y_{\mathrm{TF}}=\frac{e^{-\beta}}{4r}\left[\alpha'\left(r\beta'-r\alpha'+2\right)-2r\alpha''\right].
\end{align}
Now, by imposing the constant ${Y}_{\mathrm{TF}}=0$, the expression yields
\begin{align}\label{z24}
\alpha'\left(r\beta'-r\alpha'+2\right)-2r\alpha''=0,
\end{align}
or, alternatively
\begin{align}\label{z25}
\left[\log\alpha'-\log r+\frac{1}{2}(\alpha-\beta)\right]'=0,
\end{align}
which leads to
\begin{align}\label{z26}
\log\left(\frac{\alpha'}{r}\right)+\frac{1}{2}(\alpha-\beta)=\log \mathcal{A}_{0},
\end{align}
where $\mathcal{A}_{0}$ signifies the constant of integration.
\begin{align}\label{z27}
\alpha'e^{\alpha/2}=\mathcal{A}_{1}re^{\beta/2}, \quad \textmd{with} \quad \mathcal{A}_{1}=\textmd{constant}.
\end{align}
Then, after some manipulation, we obtain
\begin{align}\label{z28}
e^{\alpha}=\left(\mathcal{A}_{1}\int re^{\beta/2}dr+\mathcal{B}_{2}\right)^{2}.
\end{align}
Here, $\mathcal{B}_{2}$ denotes an arbitrary integration constant.

\section{Gravitational Decoupling: A Formal Framework}

The process of GD originates from a simple fluid distribution $T_{\mu\nu}$, which is extended to a more generalized form as
\begin{align}\label{o1}
T_{\mu\nu}=T^{(1)}_{\mu\nu}+T^{(2)}_{\mu\nu}+\cdot\cdot\cdot+T_{\mu\nu}^{(n)}.
\end{align}
A clear justification for this result emerges when we consider complex self-gravitational compact sources, such as those that are dissipative, electrically charged, or filled with anisotropic fluid, and apply a relatively straightforward EoS to examine the entire configuration. In this direction, understanding the significance of each fluid in a stellar configuration, and the gravitational interactions between these relativistic sources is particularly useful. By identifying the dominant source within a system, this method allows for the efficient elimination of EoS inconsistent with its properties. In the realm of GR, accomplishing this task in theory poses a significant challenge due to the nonlinear characteristics inherent in the theory. However, as the GD approach \cite{estrada2019way,gabbanelli2019causal,da2020minimal,contreras2022simple,
maurya2022gravitationally,maurya2023complexity,leon2023spherically,casadio2023axion,contreras2019general,
gabbanelli2018gravitational,hensh2019anisotropic,tello2021minimally} is specifically developed to couple or decouple the relativistic stellar formations within GR. We will observe that it becomes feasible to clarify the individual roles of each gravitational-field sector, without relying on any numerical techniques or perturbation methods. More specifically, when considering two arbitrary sources $\{M_{\mu\nu},\Theta_{\mu\nu}\}$ in Eq. (1), the contracted Bianchi identities imply $\nabla_{\mu} T^{\mu}_{\nu}+\nabla_{\mu}\Theta^{\mu}_{\nu}=0$. This situation presents two potential solutions, specifically,
\begin{align}\nonumber
M^{\mu}_{\nu}&=\nabla_{\mu}\Theta^{\mu}_{\nu}=0,
\\\nonumber
M^{\mu}_{\nu}&=-\nabla_{\mu}\Theta^{\mu}_{\nu}.
\end{align}
The initial solution suggests that each source maintains covariant conservation, thus resulting in a purely gravitational interaction between them. On the other hand, the second alternative, found to be both more intriguing and considerably more realistic, suggests an energy transfer between the field sources. Quantifying or even describing this exchange in detail would, in principle, be challenging.
Finding closed-form analytical solutions to the GR-field equations is a difficult endeavor due to their non-linear nature. Specifically, analytical solutions with particular physical relevance have only been found for certain situations \cite{stephani2009exact}. In this direction, a particularly relevant case is a spherically symmetric metric ansatz with isotropic stresses $T_{\mu\nu}$ as the gravitational source. However, in modeling more realistic scenarios, particularly those involving BHs or neutron stars, we need to couple the primary field sector (perfect fluid) with more complex forms of matter and energy. The GD scheme can lead to the solutions of the system of equations as
\begin{align}\label{z29}
T_{\mu\nu}=M_{\mu\nu}^{(\textsf{s})}+\Theta_{\mu\nu},
\end{align}
where $\Theta_{\mu\nu}$ is an additional field source whose effects on the primary source $M_{\mu\nu}^{(\textsf{s})}$ can be regulated by the deformation constant $\xi$. The coupling of the seed stress-energy tensor with extra field source makes the solution of the system highly challenging. To address this challenge, the GD scheme for decoupling the stellar distributions emerges as an effective tool. Primarily suggested for Randall-Sundrum brane-world models and generalized to explore the novel BH models, this scheme GD can be used to derive braneworld configurations from standard solutions involving perfect fluids. The GD scheme solved the system in Eq.~\eqref{z29} by constructing the solutions for the following two systems independently
\begin{align}\label{z30}
G_{\mu\nu}^{(\textsf{s})}=8\pi M_{\mu\nu}^{(\textsf{s})},
\end{align}
\begin{align}\label{z31}
G^{\ast}_{\mu\nu}=8\pi\Theta_{\mu\nu},
\end{align}
where $\{\textsl{g}^{(\textsf{s})}_{\mu\nu},M^{(\textsf{s})}_{\mu\nu}\}$ and $\{\textsl{g}^{\ast}_{\mu\nu},T^{\ast}_{\mu\nu}\}$ are obtained. Consequently, the impact of the additional field source $\Theta_{\mu\nu}$ on the source $M^{(\textsf{s})}_{\mu\nu}$ should be manifested in the measurable geometric deformation, as expressed by
\begin{align}\label{z32}
\textsl{g}_{\mu\nu}\rightarrow\textsl{g}_{\mu\nu}^{(\textsf{s})}+\textsl{g}_{\mu\nu}^{\ast}.
\end{align}
Despite the non-linearity GR relativistic equations the GD scheme allows for a straightforward combination of solutions in Eqs.~\eqref{z30} and \eqref{z31} by
\begin{align}\label{z33}
G_{\mu\nu}\equiv
G_{\mu\nu}^{(\textsf{s})}+G_{\mu\nu}^{\ast}=8\pi\left(
M_{\mu\nu}^{(\textsf{s})}+\Theta_{\mu\nu}\right)\equiv 8\pi
T_{\mu\nu},
\end{align}
where
\begin{align}\label{z34}
{M^{\mu}_{\nu}}^{(\textsf{s})}=\textmd{dia\textsl{g}}\left(\sigma^{(\textsf{s})},-P_{r}^{(\textsf{s})},
-P_{\bot}^{(\textsf{s})},-P_{\bot}^{(\textsf{s})}\right),
\end{align}
and
\begin{align}\label{z35}
\Theta^{\mu}_{\nu}=\textmd{dia\textsl{g}}\left(\Theta^{0}_{0},-\Theta^{1}_{1},
-\Theta^{2}_{2},-\Theta^{3}_{3}\right).
\end{align}
The CGD decoupling strategy modifies the metric ansatz \eqref{z7} by introducing the following linear transformations
\begin{align}\label{z36}
&\alpha\rightarrow x+\xi u,
\\\label{zz36}
&e^{-\beta}\rightarrow e^{-y}+\xi z,
\end{align}
in the temporal $\textsl{g}_{00}\equiv e^{\alpha(r)}$ and radial $\textsl{g}_{11}\equiv e^{\beta(r)}$ metric potentials, respectively. Here, the functions $\{u,z\}$ codifies the geometric deformations in the respective metric variables. These deformations are the outcomes of introducing the additional field source $\Theta_{\mu\nu}$. Interestingly, if we set $u=0$, then only $\textsl{g}_{11}$ is deformed, while $\textsl{g}_{00}$ remains unaltered. This particular deformation strategy is referred to as the MGD. However, the deformation of both the metric potentials give rise to the CGD-decoupling scheme. Thus, the modified form of the metric \eqref{z7} according CGD-decoupling approach reads
\begin{align}\label{z37}
ds^{2}=e^{(x+\xi u)}dt^{2}-(e^{-y}+\xi z)^{-1}dr^{2}-r^{2}\left(d\theta^{2}{+}\sin^{2}\theta
d\phi^{2}\right),
\end{align}
where to maintain staticity, $x$ and $y$ must be functions of the radial variable $r$ only. The gravitational-field equations for this metric can be written as
\begin{align}\label{z38}
G_{\mu\nu}=8\pi T_{\mu\nu}=8\pi\left(M_{\mu\nu}^{(\textsf{s})}+\Theta_{\mu\nu}\right),
\end{align}
whose non-zero components are defined as
\begin{align}\label{z39}
8\pi\left({M^{0}_{0}}^{(\textsf{s})}+\Theta^{0}_{0}\right)&=
\frac{1}{r^{2}}+e^{-\beta}\left(\frac{\beta'}{r}-\frac{1}{r^{2}}\right),
\\\label{z40}
8\pi\left({M^{1}_{1}}^{(\textsf{s})}+\Theta^{1}_{1}\right)&=
-\frac{1}{r^{2}}+e^{-\beta}\left(\frac{\alpha'}{r}+\frac{1}{r^{2}}\right),
\\\label{z41}
8\pi\left({M^{2}_{2}}^{(\textsf{s})}+\Theta^{2}_{2}\right)&=8\pi\left({M^{3}_{3}}^{(\textsf{s})}+\Theta^{3}_{3}\right)
=\frac{e^{-\beta}}{4}\left(2\alpha''+\alpha'^{2}
-\beta'\alpha'+2\frac{\alpha'}{r}-2\frac{\beta'}{r}\right),
\end{align}
along with the conservation equation
\begin{align}\label{z42}
(T^{1}_{1})'-\frac{\alpha'}{2}(T^{0}_{0}-T^{1}_{1})-\frac{2}{r}(T^{2}_{2}-T^{1}_{1})=0,
\end{align}
which can be split in terms of two field sources defined in \eqref{z38} as
\begin{align}\label{z43}
\left(P_{r}^{\textsf{(s)}}\right)'+\frac{\alpha'}{2}\left(\sigma^{\textsf{(s)}}+P_{r}^{\textsf{(s)}}\right)+\frac{2}{r}\left(P_{r}^{\textsf{(s)}}
-P_{\bot}^{\textsf{(s)}}\right)-\left[\left(\Theta^{1}_{1}\right)'-\frac{\alpha'}{2}\left(\Theta^{0}_{0}
-\Theta^{1}_{1}\right)-\frac{2}{r}\left(\Theta^{2}_{2}-\Theta^{1}_{1}\right)\right]=0.
\end{align}
We can now identify the effective structural quantities appearing in Eqs. \eqref{z39}--\eqref{z41} as
\begin{align}\label{z44}
&\sigma=\sigma^{\textsf{(s)}}+\Theta^{0}_{0},
\\\label{z45}
&P_{r}=P_{r}^{\textsf{(s)}}-\Theta^{1}_{1},
\\\label{z46}
&P_{\bot}=P_{\bot}^{\textsf{(s)}}-\Theta^{2}_{2}.
\end{align}
Now, by employing the linear transformations \eqref{z36} and \eqref{zz36} in the gravitational system in Eqs.~\eqref{z9}--\eqref{z11}, we can split the stellar structure equations into two systems. One characterizing the seed source governed by the stress-energy tensor, $M^{(\textsf{s})}_{\mu\nu}$
\begin{align}\label{z47}
8\pi\sigma^{\textsf{(s)}}&=
\frac{1}{r^{2}}+e^{-y}\left(\frac{y'}{r}-\frac{1}{r^{2}}\right),
\\\label{z48}
8\pi P_{r}^{\textsf{(s)}}&=
-\frac{1}{r^{2}}+e^{-y}\left(\frac{x'}{r}+\frac{1}{r^{2}}\right),
\\\label{z49}
8\pi
P_{\bot}^{\textsf{(s)}}&=\frac{e^{-y}}{4}\left(2x''+x'^{2}-x'y'+2\frac{x'}{r}-2\frac{y'}{r}\right),
\end{align}
whose dynamics is described by the following line element
\begin{align}\label{z50}
ds^{2}=e^{x(r)}dt^{2}-e^{y(r)}dr^{2}-r^{2}\left(d\theta^{2}-\sin^{2}\theta
d\phi^{2}\right),
\end{align}
where
\begin{align}\label{z51}
e^{-y(r)}\equiv 1-\frac{8\pi}{r}\int_{0}^{r}t^{2}M^{0}_{0}(t)dt,
\end{align}
is the generic expression of Misner-Sharp relativistic mass function in the absence of $\Theta_{\mu\nu}$-field source. On the other hand, the second set sourced by the $\Theta_{\mu\nu}$-field is given by
\begin{align}\label{z52}
8\pi\Theta^{0}_{0}&=
-\xi\left(\frac{z}{r^{2}}+\frac{z'}{r}\right),
\\\label{z53}
8\pi\Theta^{1}_{1}+\xi\mathcal{X}_{1}&=
-\xi z\left(\frac{1}{r^{2}}+\frac{\alpha'}{r}\right),
\\\label{z54}
8\pi\Theta^{2}_{2}+\xi\mathcal{X}_{2}&=\frac{\xi z}{r}\left(2\alpha''+\alpha'^{2}+2\frac{\alpha'}{r}\right)-\frac{\alpha z'}{4}\left(\alpha'+\frac{2}{r}\right),
\end{align}
where $\mathcal{X}_{1}=\frac{e^{-y}}{r}u'$ and
$\mathcal{X}_{2}=\frac{e^{-y}}{4}\left(2u''+\xi u'^{2}+\frac{2}{r}u'+2x'u'+y'u'\right)$. Next, the conservation relation \eqref{z43} transforms as
\begin{align}\label{z55}
-\left(P_{r}^{\textsf{(s)}}\right)'-\frac{x'}{2}\left(\sigma^{\textsf{(s)}}+P_{r}^{\textsf{(s)}}\right)-\frac{2}{r}\left(P_{r}^{\textsf{(s)}}
-P_{\bot}^{\textsf{(s)}}\right)-\frac{\xi u'}{2}\left(\sigma^{\textsf{(s)}}+P_{r}^{\textsf{(s)}}\right)+\left(\Theta^{1}_{1}\right)'-\frac{\alpha'}{2}\left(\Theta^{0}_{0}
-\Theta^{1}_{1}\right)-\frac{2}{r}\left(\Theta^{2}_{2}-\Theta^{1}_{1}\right)=0.
\end{align}
Because $G^{(\textsf{s})}_{\mu\nu}$ associated with the geometry $\{x,y\}$ satisfies the corresponding Bianchi identity, it follows that $M_{\mu\nu}^{(\textsf{s})}$ is covariantly conserved, i.e.,
\begin{align}\label{z56}
\nabla_{\mu}^{(x,y)}{M^{\mu}_{\nu}}^{(\textsf{s})}=0,
\end{align}
where $\nabla_{\mu}^{(x,y)}$ signifies the covariant differentiation with respect to the metric \eqref{z37}. In particular, we see that
\begin{align}\label{z57}
\nabla_{\mu}{M^{\mu}_{\nu}}^{(\textsf{s})}=\nabla_{\mu}^{(x,y)}{M^{\mu}_{\nu}}^{(\textsf{s})}-\frac{\xi u'}{2}\left({M^{0}_{0}}^{(\textsf{s})}-
{M^{1}_{1}}^{(\textsf{s})}\right)\delta^{1}_{\nu},
\end{align}
where $\nabla_{\mu}{M^{\mu}_{\nu}}^{(\textsf{s})}$ is determined with respect to the metric \eqref{z7}. Next, the result \eqref{z56} can be cast into the form
\begin{align}\label{z58}
\left({M^{1}_{1}}^{(\textsf{s})}\right)'-\frac{x'}{2}\left({M^{0}_{0}}^{(\textsf{s})}-{M^{1}_{1}}
^{(\textsf{s})}\right)-\frac{2}{r}\left({M^{2}_{2}}^{(\textsf{s})}-{M^{1}_{1}}^{(\textsf{s})}\right)=0.
\end{align}
This process confirms that the seed source, $M_{\mu\nu}^{(\textsf{s})}$ has been effectively separated from the gravitational system in Eqs.~\eqref{z9}--\eqref{z11}. On the other hand, by taking into account Eq. \eqref{z55} and \eqref{z56}, we obtain
\begin{align}\label{z59}
\nabla_{\mu}M^{\mu}_{\nu}=-\frac{\xi u'}{2}\left({M^{0}_{0}}^{(\textsf{s})}-
{M^{1}_{1}}^{(\textsf{s})}\right)\delta^{1}_{\nu},
\end{align}
and
\begin{align}\label{z60}
\nabla_{\mu}\Theta^{\mu}_{\nu}=-\frac{\xi u'}{2}\left({M^{0}_{0}}^{(\textsf{s})}-
{M^{1}_{1}}^{(\textsf{s})}\right)\delta^{1}_{1},
\end{align}
which contains the information regarding the exchange of stress-energy $\Delta \mathrm{E}$ between the gravitational sources, specifically
\begin{align}\label{z61}
\Delta \mathrm{E}=\frac{u'}{2}\left(\sigma^{(\textsf{s})}+P_{r}^{(\textsf{s})}\right).
\end{align}
The above expression can alternatively be defined in terms of the variables $\{x,y\}$, as
\begin{align}\label{z62}
\Delta \mathrm{E}=\frac{u'}{16\pi}\frac{e^{-y}}{r}(x'+y').
\end{align}
The explicit form of the expression \eqref{z60} reads
\begin{align}\label{z63}
\left(\Theta^{1}_{1}\right)'-\frac{\alpha'}{2}\left(\Theta^{0}_{0}
-\Theta^{1}_{1}\right)-\frac{2}{r}\left(\Theta^{2}_{2}-\Theta^{1}_{1}\right)=\frac{\xi u'}{2}\left(\sigma^{\textsf{(s)}}+P_{r}^{\textsf{(s)}}\right).
\end{align}
Hence, we deduce that the decoupling of both sources, $M_{\mu\nu}^{(\textsf{s})}$ and $\Theta_{\mu\nu}$, is achievable when there is an exchange of energy between them, as demonstrated by Eqs. \eqref{z61} and \eqref{z62}.
Next, the definition of mass function under CGD-scheme reads
\begin{align}\label{z64}
m(r)=\underbrace{4\pi\int^{r}_{0}t^{2}\sigma^{(\textsf{s})}(t)dt}_{m^{(\textsf{s})}}+
\underbrace{4\pi\xi\int^{r}_{0}t^{2}\Theta^{0}_{0}(t)dt}_{m^{(\Theta)}},
\end{align}
which implies
\begin{align}\label{z65}
m=m^{(\textsf{s})}+\xi m^{(\Theta)},
\end{align}
where $m_{s}$ and $m_{\Theta}$ has the form
\begin{align}\label{z66}
&m^{(\textsf{s})}=\frac{4\pi}{3}\sigma^{(\textsf{s})} r^{3}-\frac{4\pi}{3}\int^{r}_{0}t^{3}\sigma^{(\textsf{s})}(t)dt,
\\\label{z67}
&m^{(\Theta)}=\frac{4\pi}{3}\Theta^{0}_{0}r^{3}-\frac{4\pi}{3}\int^{r}_{0}t^{3}\Theta^{0}_{0}(t)dt.
\end{align}
Now, the combination of the Eqs. \eqref{z28}, \eqref{zz36} and \eqref{z36} produces the deformation function $u(r)$ in terms of the geometric variables $\{x,y\}$, given by
\begin{align}\label{z68}
u(r)=\frac{1}{\xi}\left[2\log\left(\mathcal{A}_{1}\int \frac{r}{\sqrt{e^{-y}+\xi z}}+\mathcal{B}_{2}\right)-x\right].
\end{align}
To solve for the function $u(r)$, we first need to calculate the deformation function $z(r)$ using a specific EoS that relates $z(r)$ and its derivatives to density. In this regard, the most suitable EoS is the one that mimics the behavior of the density, i.e., $\sigma(r)=\Theta^{0}_{0}$. Furthermore, the choice of the seed solution is also crucial, as it affects the closed-form solution for $u(r)$.
Building upon the preceding discussion, we will construct interior solutions for self-gravitating stellar configurations with anisotropic stresses, satisfying the constraint $Y_{\mathrm{TF}}$ by using the EGD-decoupling strategy. This strategy involves the following points:
\begin{itemize}
  \item Choose the best possible seed metric solutions $\{x,y\}$.
  \item Impose an EoS involving the function $z(r)$, particularly the density-like constraint $\sigma(r)=\Theta^{0}_{0}$.
  \item Evaluate $e^{-\beta}=e^{-y}+z$ by substituting the value of $z$ and replace it in Eq. \eqref{z28} to construct $e^{\alpha}$.
  \item Finally obtain $u(r)$ from Eq. \eqref{z68} by replacing the values of $e^{-y}$ and $z$.
\end{itemize}
To demonstrate this protocol, we will employ the Kohler-Chao-Tikekar metric as seed solutions in the coming sections.

\section{Completely Deformed Kohler-Chao-Tikekar Stellar solution}

This section aims to develop completely deformed stellar models featuring anisotropic stresses. To achieve this, we employ the so-called density-like constraint to derive the radial deformation function $z(r)$. This requirement provides us with
\begin{align}\label{z69}
\sigma^{(\textsf{s})}=\Theta^{0}_{0},
\end{align}
which gives rise to the differential equation (DE)
\begin{align}\label{z70}
z'+\frac{z}{r}+\frac{1}{r}+e^{-y}\left(y'-\frac{1}{r}\right)=0,
\end{align}
The provided DE allows for the determination of the radial deformation function $z$ by specifying the metric components of the seed solution. Therefore, in this work, we employ the established Kohler-Chao-Tikekar metric ansatz \cite{kohler1965zentralsymmetrische,tikekar1970suspected,andrade2023anisotropic} as the known seed solution, characterized by the following metric
\begin{align}\label{z71}
e^{x}&=H+Kr^{2},
\\\label{z72}
e^{y}&=\frac{H+2Kr^{2}}{H+Kr^{2}}.
\end{align}
Substituting the aforementioned metric components in Eq. \eqref{z70}, we achieve
\begin{align}\label{z73}
z'+\frac{z}{r}=-\frac{Hr(3H+2Kr^{2})}{(H+2Kr^{2})^{2}},
\end{align}
whose solution is given by
\begin{align}\label{z74}
z(r)=\frac{C_{0}}{r}-\frac{r^{3}K}{H+2Kr^{2}},
\end{align}
where $C_{0}$ is an arbitrary constant of integration. To formulate a non-singular stellar model maintaining the regularity constraint $z(0)=0$, we set $C_{0}=0$. We can then obtain the deformed version of the metric potential, $e^{\beta(r)}$, by substituting the value of the deformation function $z$ into Eq. \eqref{zz36}. This deformed metric potential reads
\begin{align}\label{z75}
e^{-\beta}=\frac{H+(1-\xi)Kr^{2}}{H+2Kr^{2}}.
\end{align}
Next, following the substitution of Eq. \eqref{z75} into Eq. \eqref{z28} and subsequent integration, we get the value of the modified radial metric potential $e^{\alpha(r)}$. Then, the value of the corresponding temporal deformation function $u(r)$ can now be directly determined by utilizing Eq. \eqref{z68}. The resulting values of $e^{\alpha(r)}$ and $u(r)$ along with the components of $\Theta$-sector are defined in Appendix A.
 On the other hand, the interior metric must join smoothly with the exterior metric at the boundary surface
$r=R$. This requires the continuity of both the first and second fundamental forms across the boundary surface. In the framework of GR, the Schwarzschild solution for a vacuum spacetime describes the geometric structure of the exterior geometry surrounding a spherically symmetric mass, which reads
\begin{align}\label{z81}
ds^{2}=\left(1-\frac{2M_{\textmd{Sch}}}{{\bar r}}\right)dt^{2}-\left(1-\frac{2M_{\textmd{Sch}}}{{\bar r}}\right)
^{-1}dr^{2}-r^{2}d\Omega^{2},
\end{align}
where $M_{\textmd{Sch}}$ denotes the Schwarzschild mass. Then, the following junction constraints arise form the continuity of the first and second fundamental forms,
\begin{align}\label{z82}
&\left(1-\frac{2M_{\textmd{Sch}}}{{R}}\right)=e^{\alpha(R)},
\\\label{z83}
&\left(1-\frac{2M_{\textmd{Sch}}}{{R}}\right)=e^{-\beta(R)},
\\\label{z84}
&P_{r}(R)=\left(P^{(\textsf{s})}_{r}-\xi\Theta^{1}_{1}\right)=0,
\end{align}
with
\begin{align}\label{z85}
M_{\textmd{Sch}}=m(R)=4\pi\int^{R}_{0}t^{2}\left[\sigma^{(\textsf{s})}+\xi\Theta^{0}_{0}\right]dt.
\end{align}
The conditions in Eqs.~\eqref{z82}--\eqref{z84} are both necessary and sufficient for ensuring the smooth matching of two spacetime metrics at the boundary surface $r=R$.

\section{Physical Analysis}

Compact star models are generally considered to be theoretically well-behaved if they meet essential mathematical and physical requirements. The key features of the presented stellar model are obtained by introducing the Kohler-Chao-Tikekar metric ansatz with a zero CF are particularly useful in describing the composition and development of relativistic compact configurations. The following subsections will employ this method to analyze the required criteria.

\subsection{Behavior of Physical Variables and Anisotropic Factor}

\begin{figure}
{{\includegraphics[height=2.8 in, width=3.3
in]{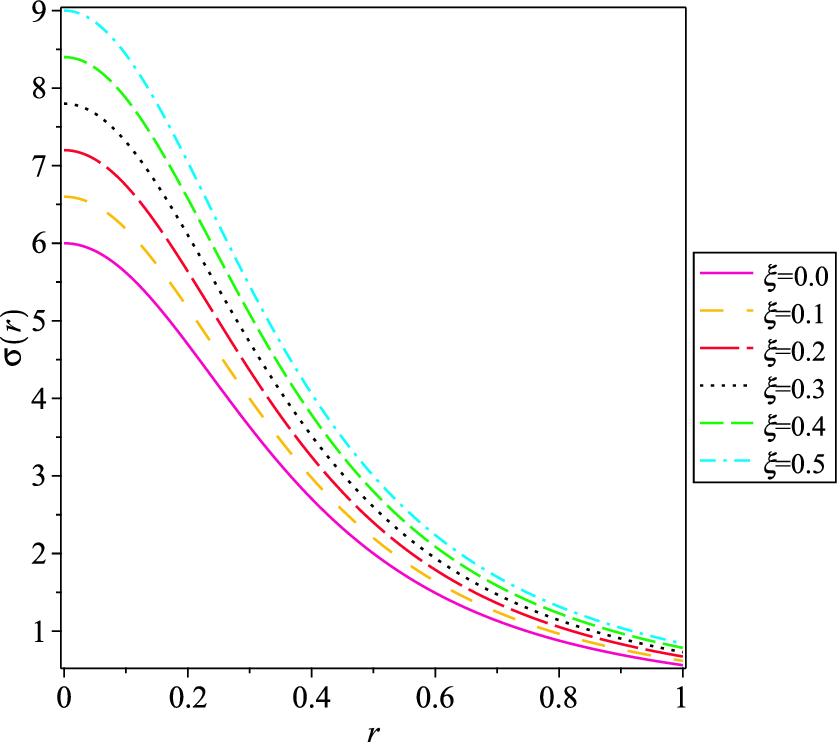}}} {{\includegraphics[height=2.8 in,
width=3.3 in]{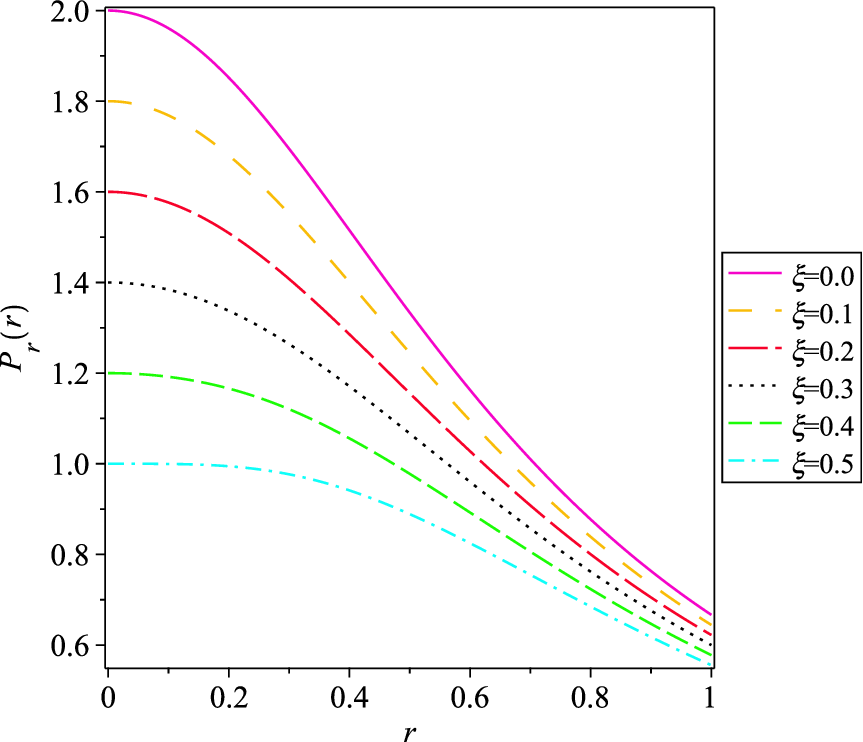}}} \caption{Plots of $\sigma$ (left panel) and $P_{r}$ (right panel) for the completely deformed Kohler--Chao--Tikekar cosmological solution as functions of the radial coordinate $r$, for different values of the deformation parameter $\xi$:
$\xi=0.0$ (solid pink),
$\xi=0.1$ (space-dashed gold),
$\xi=0.2$ (long-dashed red),
$\xi=0.3$ (solid purple),
$\xi=0.4$ (dotted black),
$\xi=0.5$ (dashed green),
and $\xi=0.6$ (dot-dashed cyan).
}\label{1f}
\end{figure}

\begin{figure}
\centering{{\includegraphics[height=2.8 in, width=3.3
in]{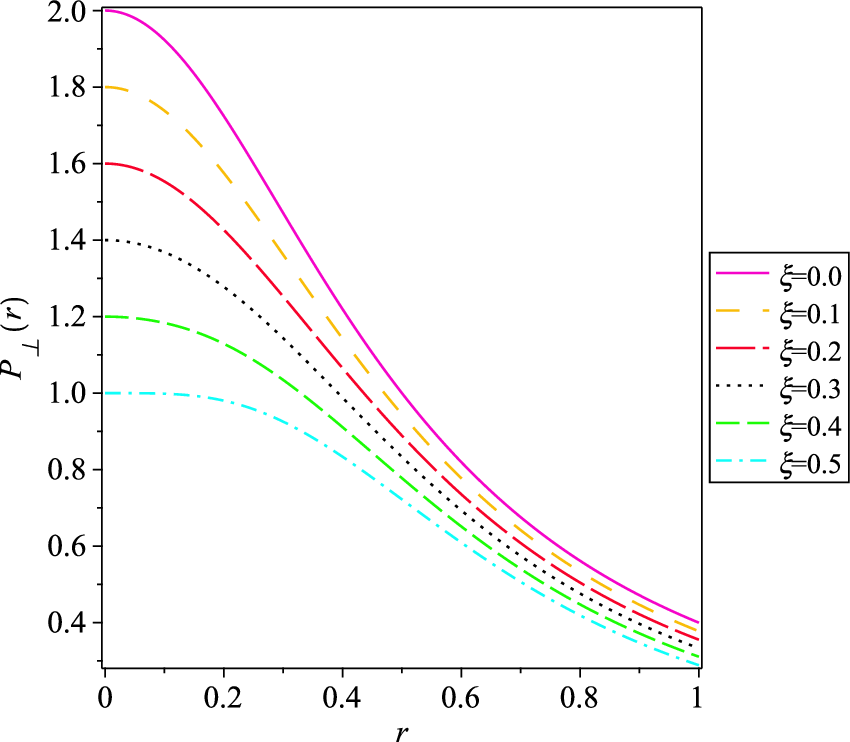}}} \centering{{\includegraphics[height=2.8 in,
width=3.3 in]{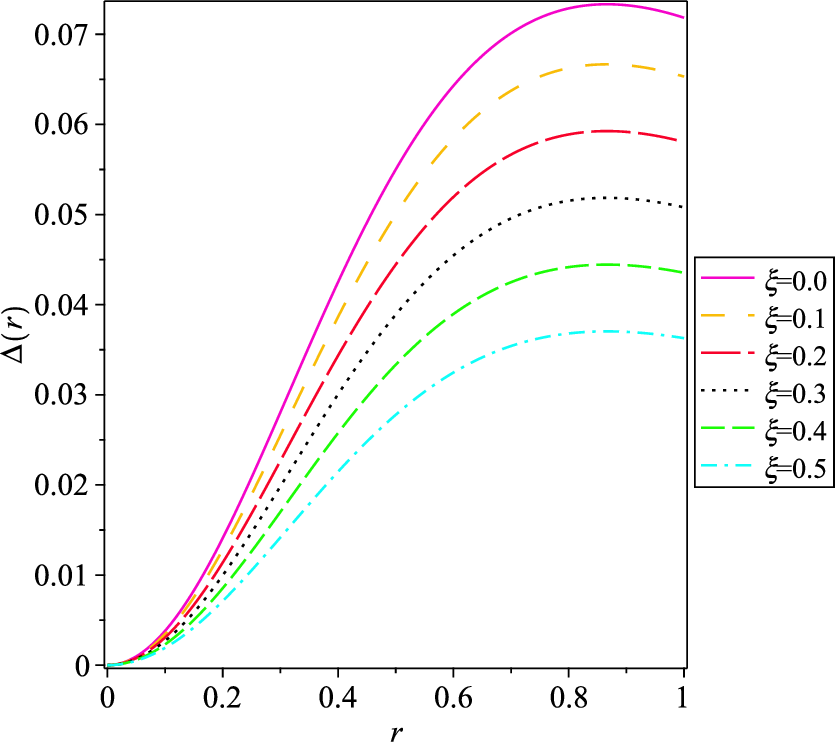}}} \caption{Plots of $P_{t}$ (left panel) and $\Delta$ (right panel) as functions of the radial coordinate $r$ for different values of the deformation parameter $\xi$:
$\xi=0.0$ (solid pink),
$\xi=0.1$ (space-dashed gold),
$\xi=0.2$ (long-dashed red),
$\xi=0.3$ (solid purple),
$\xi=0.4$ (dotted black),
$\xi=0.5$ (dashed green),
and $\xi=0.6$ (dot-dashed cyan).
}\label{2f}
\end{figure}

This subsection provides a detailed physical examination of our results based on the graphical plots displayed here. We specifically analyze the physical acceptability of the anisotropic Kohler-Chao-Tikekar stellar solution, achieved through the CGD scheme of gravitational decoupling with a zero CF constraint.
The variation of energy density corresponding to the deformation parameter $\xi$ is displayed in  Fig.~\ref{1f} (left panel). This plot determines how the density profile increases as $\xi$ increases. The density of the self-gravitational distribution as a function of $r$ increases exhibits a gradual increase as $\xi$, is varied from $0.0$ to $0.5$. Figure~\ref{1f} (right panel) illustrates the behavior of radial pressure within the self-gravitational distribution. Moving outward from the core towards the surface, the pressure decreases smoothly, ultimately reaching zero at the boundary. This vanishing pressure at the boundary indicates the absence of energy flux towards the surrounding spacetime, as expected. The impact of the decoupling constant $\xi$ shows that $P_{r}$ increases as $\xi$ decreases. This observation holds true for the $P_{\bot}$ as well, with slight variation when $\xi$ takes on small values. Furthermore, it is worth mentioning that across the entire compact fluid sphere, $P_{\bot}$ consistently exceeds its radial counterpart, with minimal deviations as shown in Fig.~\ref{2f} (left panel). The behavior of anisotropic factor $\Delta$ as a function of the coordinate, $r$, is shown in Fig.~\ref{2f} (right panel). We observe that $\Delta$ is positive at every internal location of the compact configuration. The factor $\Delta$ introduces a repulsive force $P_{r}$ exceeds $P_{\bot}$. The repulsive force counterbalances the gravitational collapse of the stellar configuration, thereby enhancing its stability. Furthermore, we highlight that the degree of anisotropy is controlled by the decoupling constant $\xi$. The value of $\Delta$ increases as the values of $\xi$ increase. Typically, we observe that as one approaches the surface layers of the self-gravitational distribution, $\Delta$ reaches its maximum.



\subsection{Energy Conditions}

\begin{figure}
\centering{{\includegraphics[height=2.8 in, width=3.3
in]{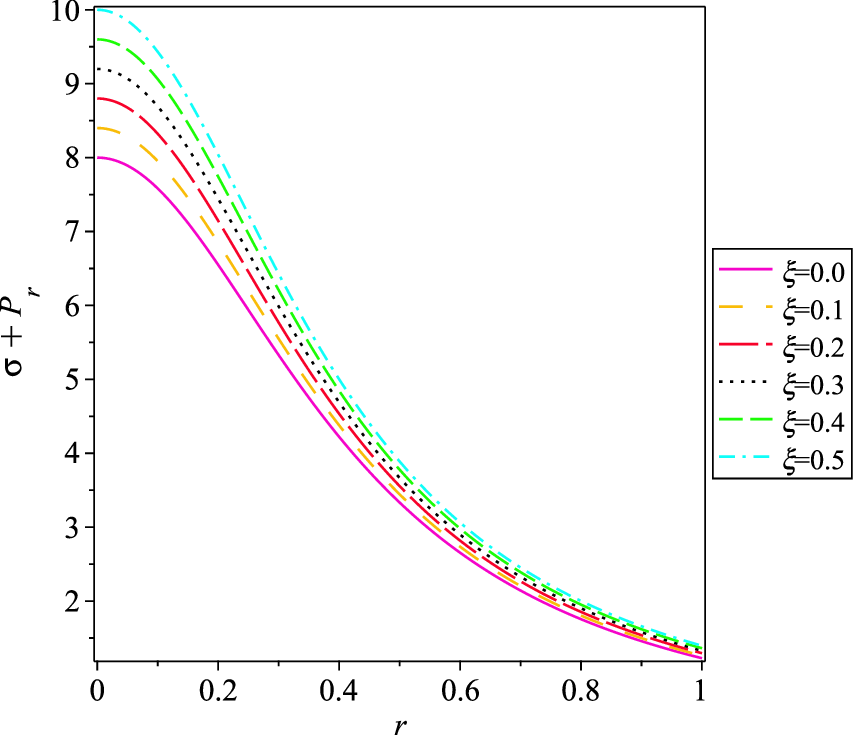}}} \centering{{\includegraphics[height=2.8 in,
width=3.3 in]{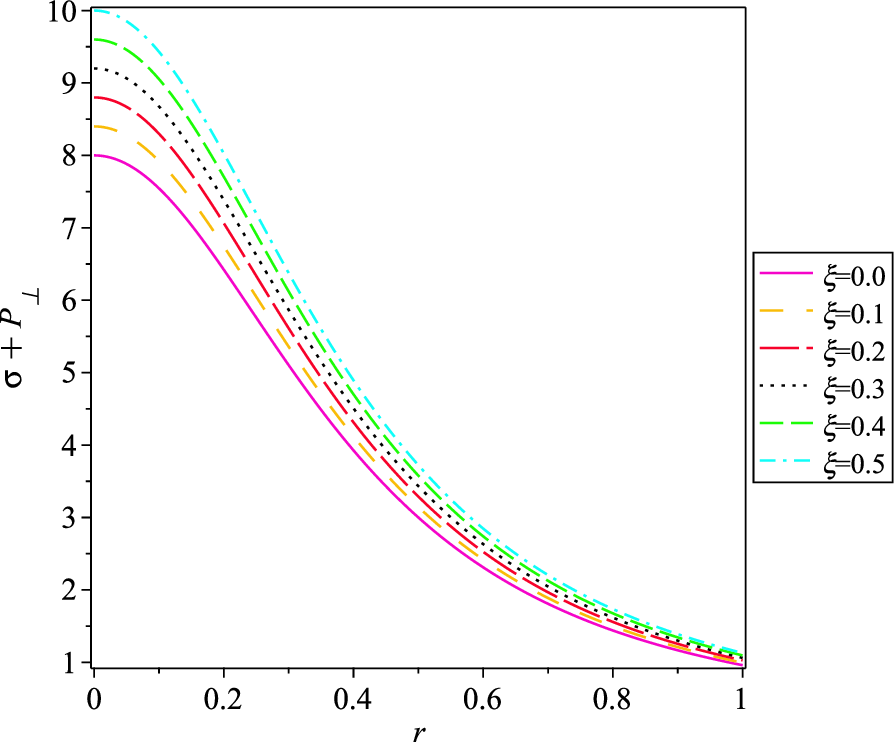}}} \caption{Plots of $\sigma+P_{r}$ (left panel) and $\sigma+P_{\bot}$ (right panel) versus $r$ for different values of the deformation parameter $\xi$. Legend is the same as in Fig.~\ref{1f}.}\label{3f}
\end{figure}

\begin{figure}
\centering{{\includegraphics[height=2.8 in, width=3.3
in]{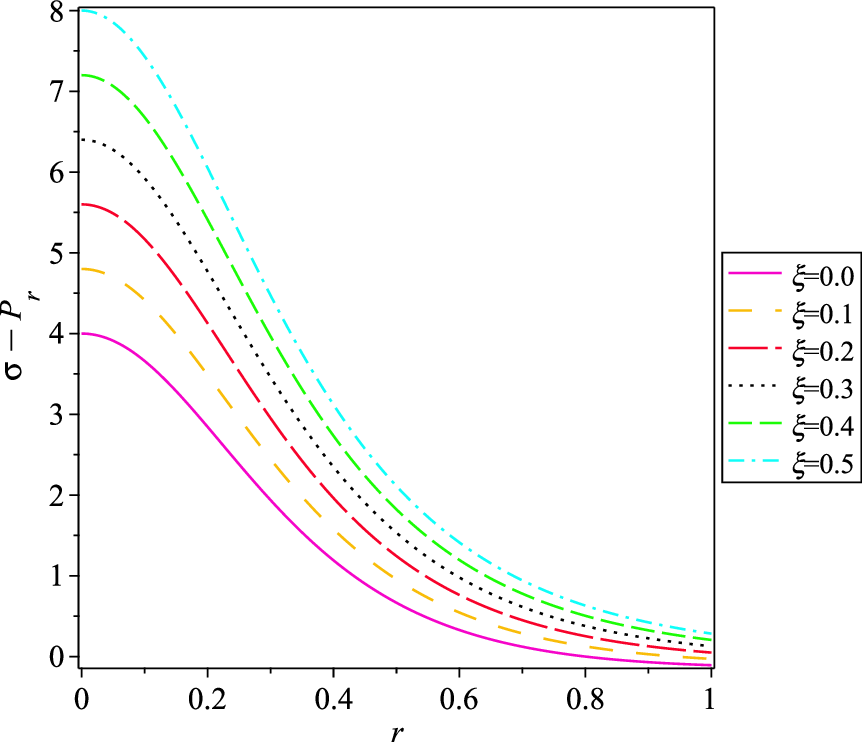}}} \centering{{\includegraphics[height=2.8 in,
width=3.3 in]{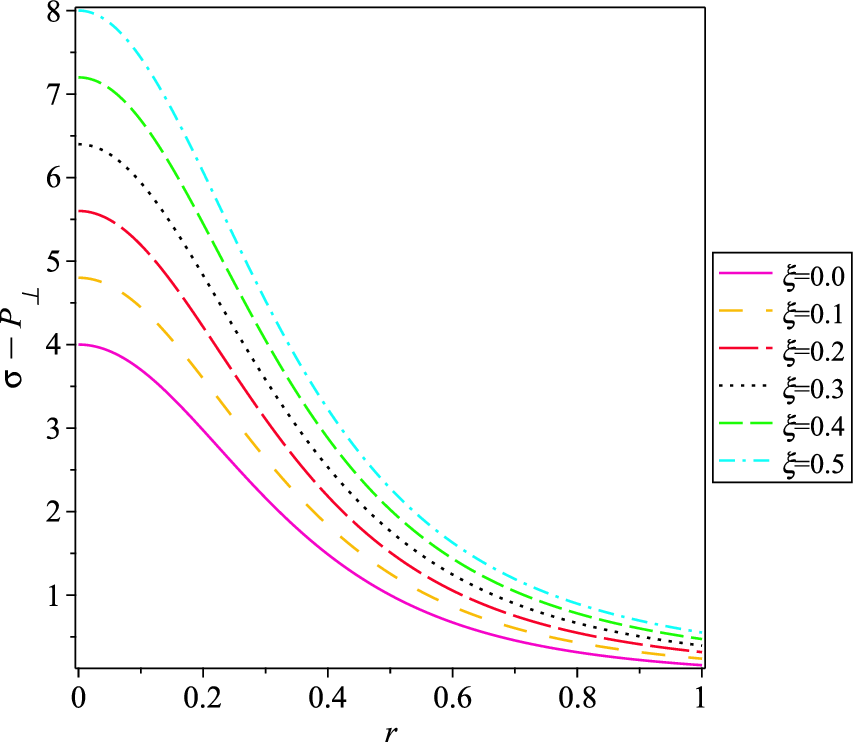}}} \caption{Plots of $\sigma-P_{r}$ (left panel) and $\sigma-P_{\bot}$ (right panel) against $r$ for various $\xi$ values subject to the Kohler-Chao-Tikekar model. Legend is the same as in Fig.~\ref{1f}.}\label{4f}
\end{figure}

\begin{figure}
\centering{{\includegraphics[height=2.8 in, width=3.3
in]{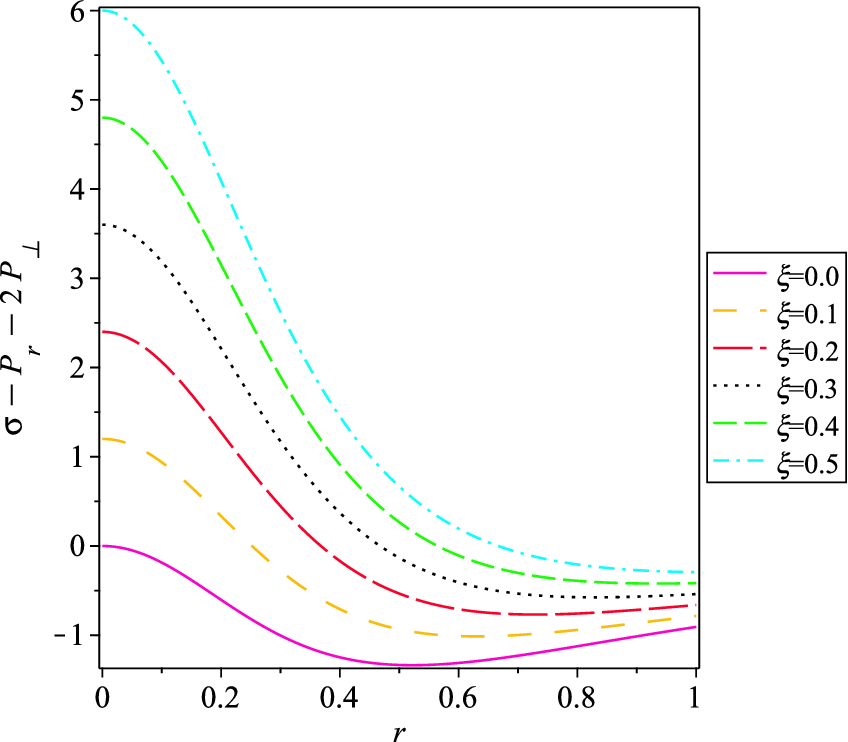}}}\caption{Plot of $\sigma+P_{r}+2P_{\bot}$ as a function of $r$ for different $\xi$ values subject to the Kohler-Chao-Tikekar model. Legend is the same as in Fig.~\ref{1f}.}\label{5f}
\end{figure}

Within a self-gravitational compact configuration, certain physical constraints, known as energy conditions, must be satisfied at every point in the system. These conditions, rooted in general relativity, relate the energy density and interior pressure of the compact star. Mathematically, these constraints are defined through the following inequalities:
\begin{description}
  \item[(i)] Weak energy condition (WEC): $\sigma>0,~\sigma+P_{r}\geq0,~\sigma+P_{\bot}\geq0$,
  \item[(ii)] Null energy condition (NEC): $\sigma+P_{r}\geq0,~\sigma+P_{\bot}\geq0$,
  \item[(iii)] Strong energy condition (SEC): $\sigma+P_{r}+2P_{\bot}\geq0$,
  \item[(iv)] Dominant energy condition (DEC): $\sigma-|P_{r}|\geq0$ and $\sigma-|P_{\bot}|\geq0$.
\end{description}
Notably, for a realistic astrophysical configuration, $\sigma$, $P_{r}$ and $P_{\bot}$ are positive, ensuring that WEC and NEC are always satisfied.
To assure that our model satisfy these constraints, we present the profiles of $\sigma+P_{r}$ and $\sigma+P_{\bot}$ in Fig.~\ref{3f}. Because both $\sigma+P_{r}$ and $\sigma+P_{\bot}$ profiles remain positive throughout the stellar interior, this confirms that the WEC and NEC is satisfied for our model. Furthermore, to validate the satisfaction of the DEC condition, we plotted the profiles of $\sigma-|P_{r}|$ and $\sigma-|P_{\bot}|$ in Fig.~\ref{4f}. Similarly, Fig.~\ref{5f} illustrates the profile of $\sigma+P_{r}+2P_{\bot}$ to verify the satisfaction of the SEC. As the plot confirms, the SEC is indeed satisfied for our model. Consequently, all the energy conditions are fulfilled throughout the dense-matter configuration.

\section{Mass and Compactness}

\begin{figure}[htbp]
\centering
\includegraphics[height=2.5in, width=3.5in]{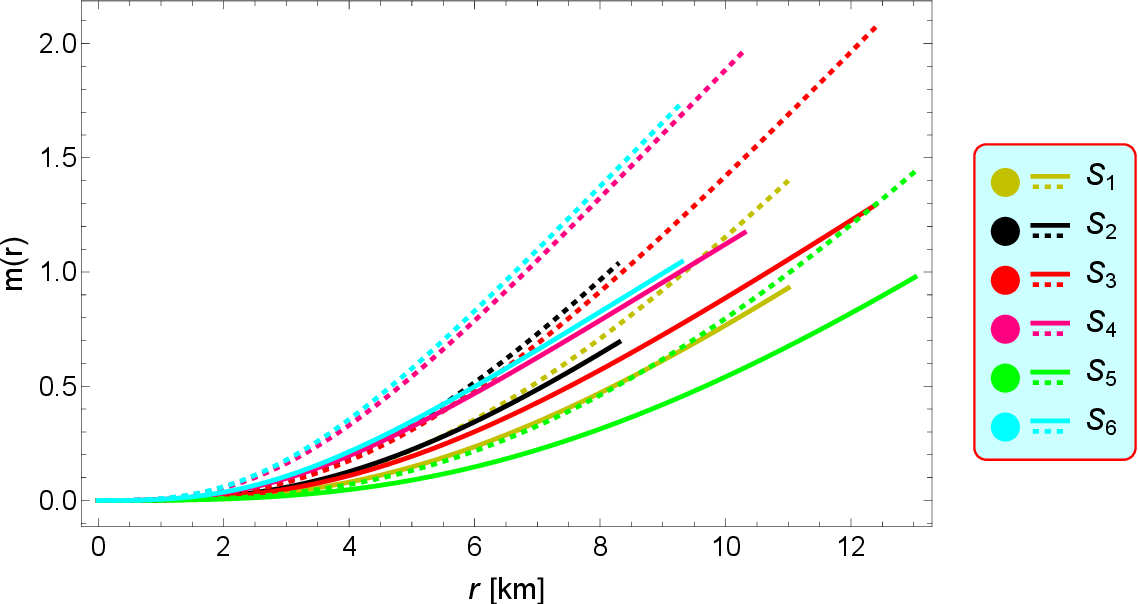}
\includegraphics[height=2.5in, width=3.5in]{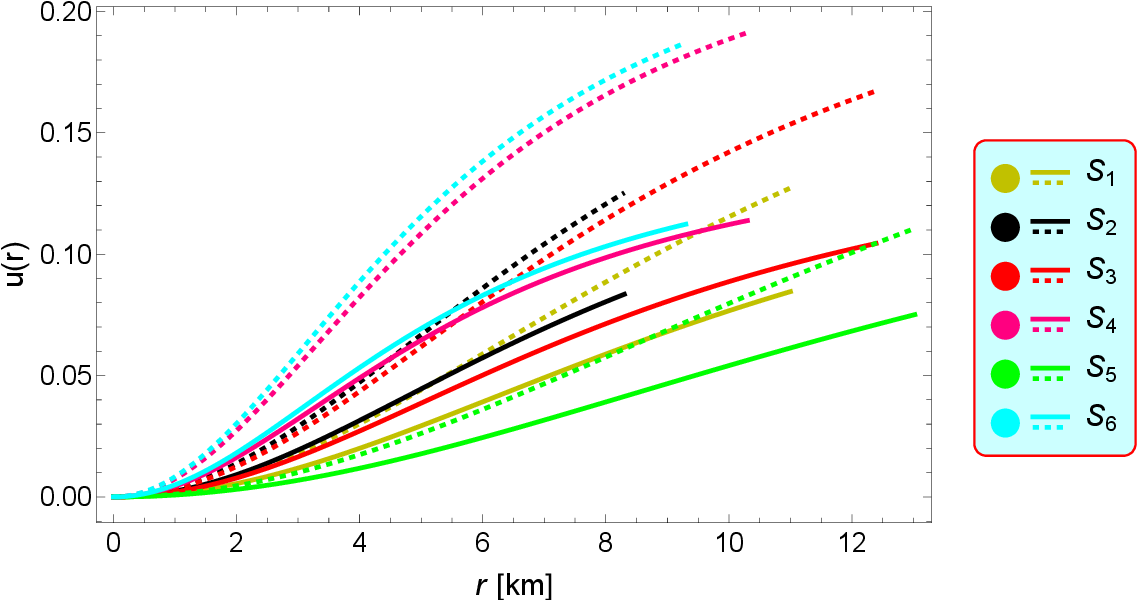}
\caption{Plots of the mass function $m(r)$ (left panel) and compactness factor $u(r)$ (right panel) versus $r$ for the compact star models \(S_i\). Dotted lines correspond to the isotropic case ($\xi = 0$), while solid lines represent the decoupled solution ($\xi \neq 0$).}
\label{w}
\end{figure}

The total mass of the presented completely deformed dense star model reads
\begin{align}\label{a1}
m(r)=4\pi \int_{0}^{r}\sigma(r)r^2dr,
\end{align}
where $\sigma=\sigma^{(\textsf{s})}+\xi \Theta^{0}_{0}$. The behavior of the total decoupled mass for our solution is shown in the left panel of Fig.~\ref{w}. It is evident that $m(r)$ increases monotonically with respect to $r$ for the different compact stars listed in Table~\ref{t}. It can be observed that the predicted masses and radii remain within the viable limits of compact stars.

The compactness factor is defined as
\begin{align}\label{a2}
u(r)=\frac{m(r)}{r}.
\end{align}
The graphical analysis of $u(r)$ for the considered stars is displayed in the right panel of Fig. \ref{w}. The profile of $u(r)$ remains well within the Buchdahl limit \cite{buchdahl1959general}, which states that for any $(3+1)$-dimensional stable matter configuration,
$u < \frac{4}{9}.$
\begin{table}[htbp]
\centering
\caption{Observed compact stars used for comparison, labelled as \(S_i\). Masses are in units of \(M_\odot\) and radii in km.}
\label{t}
\begin{tabular}{c l c c}
\hline\hline
Label & Source (name) & Mass [$M_\odot$] & Radius [km] \\
\hline
$S_{1}$ & EXO 1745--248      & 1.40   & 11.00  \\
$S_{2}$ & LMC X--4           & 1.04   & 8.301  \\
$S_{3}$ & PSR J0740+6620     & 2.072  & 12.39  \\
$S_{4}$ & PSR J1614--2230    & 1.97   & 10.30  \\
$S_{5}$ & PSR J0030+0451     & 1.44   & 13.02  \\
$S_{6}$ & 4U 1608--52        & 1.74   & 9.30   \\
\hline\hline
\end{tabular}
\end{table}

\subsection{Energy Exchange}

\begin{figure}
\centering{{\includegraphics[height=2.8 in, width=3.3
in]{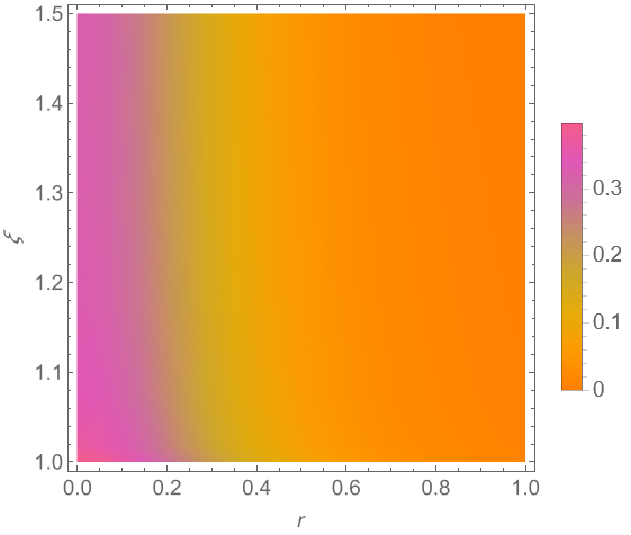}}} \centering{{\includegraphics[height=2.8 in,
width=3.3 in]{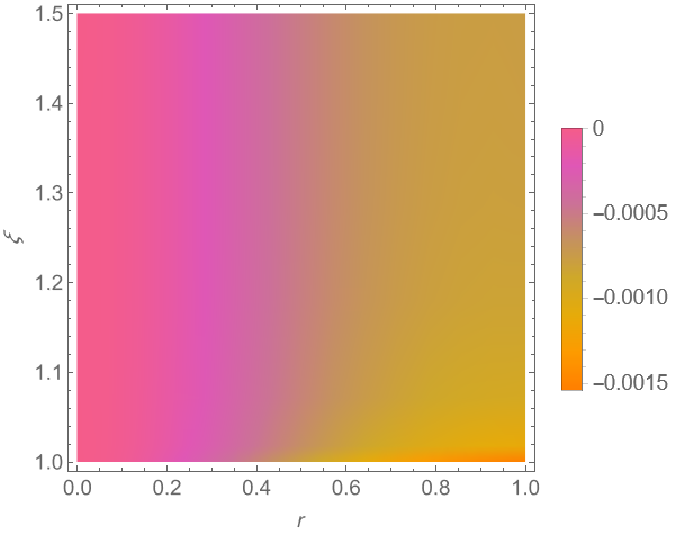}}} \caption{Left panel: Density plot of exchange of energy for $\mathcal{A}_{1}=0.5$ and $\mathcal{B}_{1}=1.0$: Density plot of exchange of energy for $\mathcal{A}_{1}=0.3$ and $\mathcal{B}_{1}=0.1$.}\label{6f}
\end{figure}

This subsection will explore a fascinating structural aspect of the solution: the energy exchange between the seed gravitational-field source $M_{\mu\nu}^{(\textsf{s})}$ and the unknown field component $\Theta_{\mu\nu}$. In this respect, the expression for the energy exchange is defined as
\begin{align}\nonumber
\Delta \mathrm{E}=\frac{A}{B},
\end{align}
where
\begin{align}\nonumber
A=&\left[K\left(H+(-3+r^{2})K\right)\left\{2(\xi-1)(H+2Kr^{2})\sqrt{H-(\xi-1)Kr^{2}}
\left((2\xi-1)H+(\xi-1)Kr^{2}\right)-\sqrt{2}(1+\xi)H\right.\right.
\\\label{k87}
\times&\left.\left.\left(-H+(\xi-1)Kr^{2}\right)\arctan\left(\frac{\sqrt{2}\sqrt{H-(\xi-1)Kr^{2}}}
{\sqrt{\xi-1}\sqrt{H+2Kr^{2}}}\right)\right\}\right],
\end{align}
and
\begin{align}\nonumber
B=&\left[-16(\xi-1)\xi(H+2Kr^{2})^{3}\left(
H-(\xi-1)Kr^{2}\right)^{3/2}\right.
-\left.8\sqrt{2}\pi\sqrt{\xi-1}\xi(\xi+1)H(H+2r^{2}K)^{5/2}\left(H-K(\xi-1)r^{2}\right)\right.
\\\label{z87}
\times&\left.\arctan\left(\frac{\sqrt{2}\sqrt{H-(\xi-1)Kr^{2}}}
{\sqrt{\xi-1}\sqrt{H+2Kr^{2}}}\right)\right].
\end{align}
Particularly, the sign of the quantity $\Delta \mathrm{E}$ determines the nature of the energy transfer between the gravitational-field sources:
\begin{description}
                                          \item[(i)] A positive of $\Delta \mathrm{E}$ signifies that the generic-field source $\Theta_{\mu\nu}$ transfers energy to its surroundings.
                                          \item[(ii)] A negative of $\Delta \mathrm{E}$ signifies that the seed-gravitational source $M_{\mu\nu}^{(\textsf{s})}$ gains energy to from its surroundings.
                                        \end{description}
This energy exchange transition is illustrated in Fig.~\ref{6f} on the $\xi-r$ plane. The left panel of density diagram (Fig.~\ref{6f}) describes that $\Delta \mathrm{E}>0$ on $\xi-r$ plane within the slef-gravitational distribution when $r\in(0,1)$ and it is maximum within the compact configuration at $\xi=1.5$. This indicates that the seed-gravitational source $M_{\mu\nu}^{(\textsf{s})}$ gains energy from the surrounding, whereas there is no energy exchange between the sources at the center and boundary of the stellar configuration. In other words the generic-field source $\Theta_{\mu\nu}$ is giving the maximum energy for the deformation parameter $\xi\approx1.5$. The density diagram indicates a significant transfer of energy occurring in the core region for $\xi<1.1$. The left panel of the density diagram reveals that $\Delta \mathrm{E}<0$ on the $\xi-r$ plane within the object, specifically when $r$ lies between $0$ and $1$. This suggests that the extra-field source $\Theta_{\mu\nu}$ absorbs energy from the environment in the context of zero CF.

\section{Conclusions}

In this article, we have detailed the GD approach within the formalism of CGD decoupling to construct anisotropic stellar configurations by choosing the zero CF condition as an EoS. Our results confirm that GD through the CGD scheme offers an effective framework for constructing generic solutions to the GR field equations. This is achieved by employing the density-like constraint and the $Y_{\mathrm{TF}}=0$ condition. By employing the CGD decoupling approach, we have comprehensively examined the most generic method for decoupling two gravitational-field sources $\{M_{\mu\nu},\Theta_{\mu\nu}\}$ in GR. This method is applied to two scenarios: firstly, when both the metric variables $\{\textsl{g}_{00},\textsl{g}_{11}\}$ are deformed, secondly, when it remains valid throughout all spacetime regions, regardless of matter presence.

The above-mentioned result was established by a thorough and precise analysis of Bianchi's identities. This examination shows that a successful decoupling of the GR-field Eqs.~\eqref{z39}--\eqref{z41} can only be achieved through an energy exchange between both stellar-field sources, as depicted in Eqs.~\eqref{z59} and \eqref{z60}. The decoupling of both the stellar-field sectors results in two independent gravitational systems: the already known seed solution associated with $M_{\mu\nu}$, defined in Eqs. \eqref{z47}--\eqref{z49}, and an extra unknown field source $\Theta_{\mu\nu}$ that corresponds to the quasi-Einstein stellar system, presented in Eqs.~\eqref{z52}--\eqref{z54}. Each gravitational-field system is accompanied by its corresponding conservation equation, provided in Eqs. \eqref{z58} and \eqref{z63}, respectively. Particularly, we have employed the CGD scheme, which involves deforming both the metric potentials, i.e., radial as well as temporal, by introducing linear transformations defined in Eqs.~\eqref{z36}--\eqref{zz36}. With the help of this scheme, we have successfully formulated the complexity-free Kohler-Chao-Tikekar model within the anisotropic domain. The obtained stellar solution characterizes a self-gravitational sphere with anisotropic stresses. In other words, by employing the CGD decoupling scheme, we have successfully derived an anisotropic generalization of a stellar solution. This model satisfies the basic requirements for a realistic description of a stellar interior. In other words, both the metric and the structural variables associated with the material sector (such as $\sigma$, $P_{r}$, and $P_{\bot}$) in this solution are finite and exhibit regular behavior as functions of $r$. Additionally, the function $e^{\alpha(r)}$ is a monotonically increasing function of $r$, while the functions $\left\{e^{-\beta(r)},\sigma, P_{r}, P_{\bot}\right\}$ are monotonically decreasing functions of $r$. Within the stellar compact object, all relevant physical quantities of the matter sector achieve their maxima precisely at the center. Additionally, we also ensured that the anisotropic factor $\Delta>0$ is throughout the self-gravitational stellar configuration object, except at the center where it vanishes, which is crucial for the stability of this compact structure. Our model satisfies all the necessary physical conditions (WEC, DEC, SEC) for a realistic description of matter behavior. Furthermore, it is important to note that the approach employed in this investigation provides another viable tool for obtaining interior solutions within the domain of pressure, anisotropy which is characterized by zero CF. This approach could be particularly interesting from the perspective of exploring the effects of decoupling on the complexity of anisotropic fluid spheres. Furthermore, it would be fascinating to extend this approach to other levels of the CF.

Looking ahead, our approach can be extended to more general stellar models, including rotating or charged configurations and alternative equations of state. The role of the deformation parameter as a descriptor of energy exchange also opens prospects for applications in modified gravity and dark matter environments, providing a broader astrophysical scope for the gravitational decoupling scheme. Moreover, our framework may also be employed to analyze the effects of nonmetricity- and torsion-based \cite{mahmoud2025anisotropic} gravity models on the gravitational decoupling of astrophysical systems.


\vspace{0.3cm}


\noindent {\bf Data Availability Statement:} This manuscript has no associated data or the data will not be deposited.

\vspace{0.5cm}

\noindent {\bf Declaration of Competing Interest:}
The authors declare that they have no known competing financial interests or personal relationships that could have appeared to influence the work reported in this paper.

\begin{acknowledgments}
The work of KB was supported in part by the JSPS KAKENHI Grant Numbers 21K03547, 24KF0100, 25KF0176 and Competitive Research Funds for Fukushima University Faculty (25RK011).
\end{acknowledgments}

\section*{Appendix A}

In this Appendix, we describe some important expressions that would help understand our mathematical computations.

The value of the modified temporal metric function obtained as a result of CGD decoupling scheme can be defined as
\begin{align}\label{z76}
e^{\alpha/2}=\mathcal{A}_{1}\left[\frac{2\mathcal{X}(r)\left((\xi-1)r^{2}K-H\right)-
\sqrt{2}\left(1+\xi\right)H\sqrt{H-(\xi-1)Kr^{2}}\arctan\left(\frac{\sqrt{2H-2(\xi-1)Kr^{2}}}
{\mathcal{X}(r)}\right)}{4(\xi-1)^{1/2}K\mathcal{X}(r)\sqrt{\frac{H-(\xi-1)Kr^{2}}
{H+2Kr^{2}}}}\right]
+\mathcal{B}_{1},
\end{align}
where the auxiliary function $\mathcal{X}(r)$ is defined as
\begin{align}\label{x1}
\mathcal{X}(r)=&\sqrt{(\xi-1)(H+2Kr^{2})}.
\end{align}
Then, the value of the temporal deformation function $u(r)$ can be attained with the help of modified metric variable $e^{\alpha}$, which is given by
\begin{align}\nonumber
u(r)=&\frac{1}{\xi}\left[2\log\left\{\mathcal{A}_{1}\left(\frac{2\mathcal{X}(r)\left((\xi-1)r^{2}K-H\right)-
\sqrt{2}\left(1+\xi\right)H\sqrt{H-(\xi-1)Kr^{2}}\arctan\left(\frac{\sqrt{2H-2(\xi-1)Kr^{2}}}
{\mathcal{X}(r)}\right)}{4(\xi-1)^{1/2}K\mathcal{X}(r)\sqrt{\frac{H-(\xi-1)Kr^{2}}
{H+2Kr^{2}}}}\right)\right.\right.
\\\label{z78}
+&\left.\left.\mathcal{B}_{1}\right\}+\mathcal{B}_{1}-\log(H+Kr^{2})\right].
\end{align}
The function $u(r)$ is responsible for measuring the effects CDG decoupling on the self-gravitational source. The values of the structural variables associated with the $\Theta$-sector reads
\begin{align}\label{z79}
\Theta^{0}_{0}=&\frac{K(3H+2Kr^{2})}{(H+2Kr^{2})^{2}},
\\\nonumber
\Theta^{1}_{1}=&\left[\phi_{1}\left\{HK\left(-4+2(3+2r)\xi-r\xi^{2}\right)r^{2}+\left(H^{2}
\left(-2+(4+r)\xi\right)\right)-r^{4}(\xi-1)(-2+3r\xi)K^{2}\right\}+K\left((2+r\xi)H\right.\right.
\\\label{z80}
+&\left.\left.r^{2}(2+3r\xi)K\right)\phi_{2}\right]/\left[\xi(H+Kr^{2})(H+2Kr^{2})
\left\{2\sqrt{\xi-1}\sqrt{H+2Kr^{2}}\left(H-(\xi-1)Kr^{2}\right)+\phi_{2}\right\}\right],
\\\nonumber
\Theta^{2}_{2}=&\frac{2K^{2}(2H+Kr^{2})+K(H+Kr^{2})(3H+2Kr^{2})}{2(H+Kr^{2})^{2}(H+2Kr^{2})}-\frac{HK^{2}r^{2}\left\{-\phi_{1}\left((2\xi-1)H+(\xi-1)Kr^{2}\right)
+\phi_{2}\right\}}{\xi\left(H+Kr^{2}\right)(H+2Kr^{2})^{2}\left\{\phi_{1}(H-(\xi-1)Kr^{2})+\phi_{2}\right\}}
\\\nonumber
-&\left[rHK(\xi-1)^{9/2}(\xi+1)(H+Kr^{2})(H+2Kr^{2})^{2}\left(H-(\xi-1)Kr^{2}\right)^{2}
\left\{2\mathcal{X}(r)\left((\xi-1)Kr^{2}-H\right)-\phi_{2}\right\}\right.
\\\nonumber
\times&\left.\left\{2\mathcal{X}(r)\left((2\xi-1)H+(\xi-1)Kr^{2}\right)+\phi_{2}\right\}+
r^{3}HK^{2}(\xi-1)^{9/2}(\xi+1)(H+Kr^{2})\left(H-(\xi-1)Kr^{2}\right)^{2}\right.
\\\nonumber
\times&\left.(H+2Kr^{2})^{2}\left\{2\mathcal{X}(r)\left((2\xi-1)H+(\xi-1)Kr^{2}\right)+\phi_{2}\right\}
^{2}-rHK(1+\xi)(\xi-1)^{2}\left\{-(H+Kr^{2})^{2}(H+2Kr^{2})^{3/2}\right.\right.
\\\nonumber
\times&\left.\left.4\left(H^{3}-H^{2}Kr^{2}(\xi-5)-6HK^{2}r^{4}+2K^{3}(\xi-1)^{2}r^{6}\right)(\xi-1)^{4}
\left\{-2\mathcal{X}(r)\left((\xi-1)Kr^{2}-H\right)+\phi_{2}\right\}+4HKr^{2}\right.\right.
\\\nonumber
\times&\left.\left.(H+Kr^{2})^{2}(H+2Kr^{2})^{3/2}\left(H-(\xi-1)Kr^{2}\right)(\xi-1)^{4}(\xi+1)
\left\{-2\mathcal{X}(r)\left((\xi-1)Kr^{2}-H\right)+\phi_{2}\right\}+2Kr^{2}\right.\right.
\\\nonumber
\times&\left.\left.(H+2Kr^{2})^{2}\left(H-(\xi-1)Kr^{2}\right)^{2}(\xi-1)^{5/2}\left\{-2\mathcal{X}(r)
\left((\xi-1)Kr^{2}-H\right)+\phi_{2}\right\}^{2}-(H+Kr^{2})(H+2Kr^{2})^{2}\right.\right.
\\\nonumber
\times&\left.\left.\left(H-(\xi-1)Kr^{2}\right)^{2}(\xi-1)^{5/2}\left\{-2\mathcal{X}(r)
\left((\xi-1)Kr^{2}-H\right)+\phi_{2}\right\}^{2}+8K(H+Kr^{2})^{2}r^{2}\left\{-2(H+2Kr^{2})^{2}
\right.\right.\right.
\\\nonumber
\times&\left.\left.\left.\left(H-K(\xi-1)r^{2}\right)^{3}(\xi-1)^{9/2}-\sqrt{2}H
(H+2Kr^{2})^{3/2}\left(H-(\xi-1)Kr^{2}\right)^{5/2}(\xi-1)^{4}+(\xi+1)\right.\right.\right.
\\\nonumber
\times&\left.\left.\left.\arctan\left(\frac{\sqrt{2}\sqrt{H-(\xi)Kr^{2}}}
{\sqrt{\xi-1}\sqrt{H+2Kr^{2}}}\right)\right\}+4K(H+Kr^{2})^{2}r^{2}\left\{-6(H+2Kr^{2})^{3}(\xi-1)^{11/3}
-\sqrt{2}H(H+2Kr^{2})^{5/2}\right.\right.\right.
\\\nonumber
\times&\left.\left.\left.(H-K(\xi-1)^{3/2}r^{2})(\xi-1)^{5}(\xi+1)\arctan\left(\frac{\sqrt{2}\sqrt{H-(\xi)Kr^{2}}}
{\sqrt{\xi-1}\sqrt{H+2Kr^{2}}}\right)\right\}\right\}+HKr^{2}(H-(\xi-1)Kr^{2})^{2}\right.
\\\nonumber
\times&\left.(H+2Kr^{2})^{2}(\xi-1)^{9/2}(1+\xi)\left\{2\mathcal{X}(r)
\left((\xi-1)Kr^{2}-H\right)-\phi_{2}\right\}\left\{-2\mathcal{X}(r)
\left((\xi-1)Kr^{2}-H\right)+\phi_{2}\right\}\left\{\mathcal{B}_{1}H\right.\right.
\\\nonumber
-&\left.\left.2rK+r^{2}\mathcal{B}_{1}K-(H+Kr^{2})\log(H+Kr^{2})+2(H+Kr^{2})\right\}\log\left\{\frac{\left\{-2\mathcal{X}(r)
\left((\xi-1)Kr^{2}-H\right)-\phi_{2}\right\}}{4K\sqrt{H+2Kr^{2}}\sqrt{\frac{H-K(\xi-1)r^{2}}{H+2Kr^{2}}}
(\xi-1)^{3/2}}\right\}\right]
\\\label{s80}
&/\left[rH(H+Kr^{2})(H+2Kr^{2})^{3}\left(H-K(\xi-1)r^{2}\right)^{2}(\xi-1)^{9/2}\xi(\xi+1)
\left\{-2\mathcal{X}(r)
\left((\xi-1)Kr^{2}-H\right)+\phi_{2}\right\}^{2}\right],
\end{align}
where $\phi_{1}$ and $\phi_{2}$ are auxiliary functions defined as
\begin{align}\label{x122}
\phi_{1}=&-2K\sqrt{\xi-1}\sqrt{H+2Kr^{2}},
\\\label{x3}
\phi_{2}=&\sqrt{2}(1+\xi)H\sqrt{H-(\xi-1)Kr^{2}}\arctan\left(\frac{\sqrt{2}\sqrt{H-(\xi-1)Kr^{2}}}
{\sqrt{\xi-1}\sqrt{H+2Kr^{2}}}\right).
\end{align}

\vspace{0.3cm}

\providecommand{\noopsort}[1]{}\providecommand{\singleletter}[1]{#1}%

\end{document}